# SECURE ELECTRONIC LOCK USING PIC 16F628A MICROCONTROLLER

Muhanad Hayder Mohammed

*Lecturer in Computer Science Department, University of Kerbala Iraq*
*muhanad.hayder@uokerbala.edu.iq*

***Abstract:*** *The proposed system implements an electronic embedded lock that provides a great benefit over a traditional lock, which uses only a manual key. If in case, the key is lost or stolen then anyone can open the lock using the key. On the other hand, losing a long and complex password or getting it stolen is harder as compared to a traditional key. Further, a combination of both, the manual key with a computerized password makes the system more secure. A longer password will reduce the possibilities of getting the code broken and opening the lock. Our system comprises of a keypad and HD44780 20x2 LCD along with a PIC16f628a microcontroller. The firmware controls these components in such a way that interaction with keypad is very easy and smooth. The LCD provides user with messages and notifications to inform about the current system state. User can perform operations such as opening and closing the lock, changing the current password in the microcontroller EEPROM and clearing a single digit while entering the password when wrong digit entered (back space). The proposed system's firmware is developed using assembly language with MPLAB development environment. It is tested and implemented at the actual hardware level with proper functioning of the system and it is completely bug free.*

## I. INTRODUCTION

Today, we are living in the in the of embedded systems surrounded by devices that based on the embedded systems like cars, washing machine microwave-oven, medical equipment's etc.

An embedded system is a computer system designed for specific control functions often with real-time computing constraints. It is embedded as part of a complete device which often including hardware and mechanical parts. By contrast, a general-purpose computer, such as a personal computer (PC), is designed to be flexible to meet a wide range of end-user needs. Embedded systems control many devices in common use today [1].

One of prominent example of an embedded system is a microcontroller, which is a small and tiny computer designated to perform some specific tasks. A Microcontroller program (firmware) is the one, which decides what functionality the microcontroller provides to a user. A program that has the ability to run on a microcontroller without the need of an operating system is called as a firmware. That means, a firmware has the privilege to access the hardware directly. This paper tries to explain an electronic lock firmware in detail. The basic idea of microcontroller is to collect all the input and output peripherals in one simple circuit, which represent the microcontroller instead of the large and sophisticated computer with microprocessor and large numbers of peripherals [1].

The firmware directly deals with peripherals and Input / Output ports to give complete functionality of microcontroller.

## II. MICROPROCESSOR vs MICROCONTROLLER

Microprocessor differs from a microcontroller in many aspects. First and the most important aspect is its architecture. In order for a microprocessor to function properly, other components such as memory, peripherals and input output ports must be connected to it. In short, we can say that a microprocessor is the heart of the computer and it works a group with other peripherals / parts of the computer system. On the other hand, a microcontroller is designed to be comprised as a single unit which can perform independently. No other external component is needed for its application because all necessary peripherals and ports are already built into it. Ultimately, it saves the time and space needed to construct devices [2].

## III. SYSTEM IMPLEMENTATION

The system is implemented using assembly language. The purpose of using this language is to get a better picture and understanding of the PIC architecture. Another positive aspect of using assembly language is that it gives very wide and flexible way to interact with microcontrollers even though its implementation is harder than BASIC or C languages, which are also used as a language for programming microcontrollers. There are mainly two parts of the system implementations:

## A. Keypad Implementation

The system dedicates PORTB for keypad implementation in which 4 pins are reserved for columns and the remaining 4 pins for rows. Hence, the system uses these 8 pins for construction of 16 keys matrix, which makes an efficient use of the limited ports provided in PIC16F628A. Following is the Table (1) PORTB pins assignments for keypad.

*Table1: PORTB pins assignments for keypad*

|  | RB4(column0) | RB5(column 1) | RB6(column 2) | RB7(column 3) |
|---|---|---|---|---|
| RB0(row0) | 1 | 2 | 3 | A =Back space |
| RB1(row1) | 4 | 5 | 6 | B= lock the system |
| RB2(row2) | 7 | 8 | 9 | C= Modify pass. |
| RB3(row3) | * | 0 | # | D=Enter |

There is an important precaution which must be taken while implementing the keypad because, the keypad processing rely on a mechanical process (pressing and releasing keys). This phenomenon generates a spark, which influences the electrical properties of pin while pressing and releasing the key, which causes the pin status to be unstable and can't be recognized properly whether it is 0 or 1. So, to overcome this, a delay must be provided to give enough time for the pin status to be stable and can be read correctly, this time could be 20ms or more.

1. Key Scanning

The keypad design is influenced by electrical phenomena, which states that, if we short circuit two pins, one with zero and the other with high voltage, then both the pin voltage would drop to zero. This fact could be used for designing the keypad.

As mentioned above, PORTB is reserved for the keypad and the most significant 4 bits are used for column indexing, which are set as input port (for input port, we must set TRISB register to high [3]) to check whether they have dropped to 0 voltage when they are attached with the selected row after pressing the key.

On the other hand, the least significant 4 bits of PORTB are set to output, and are used for row indexing. Since they are the output ports, they can be set to high (1) or low (0) by the firmware. By setting each row to zero for every key scan, the system can check each column, to see if any one of them is set to 0. If so, this means that one key of that row has been pressed. This process is repeated for all rows to check all the keys. The functions responsible for scanning key are *row_scan* and *col_scan*.

1.1 Row scan

This function sets every row to zero at a time and calls *col_scan* function to check if any column pins drop to zero voltage. This function then increments the *key* variable for each row assignments, to index the row.

1.2 Col_scan

This function scans every column pin to check if any one of them is set to zero, if so then it calls *find_key* function and finds the key value throw the index constructed using the variable *key* and the hard coded value getting from working register *w* throw *col_scan* Function. Figure (1) show the keypad structure connected to PIC 16F628A microcontroller and LCD pins assignments.

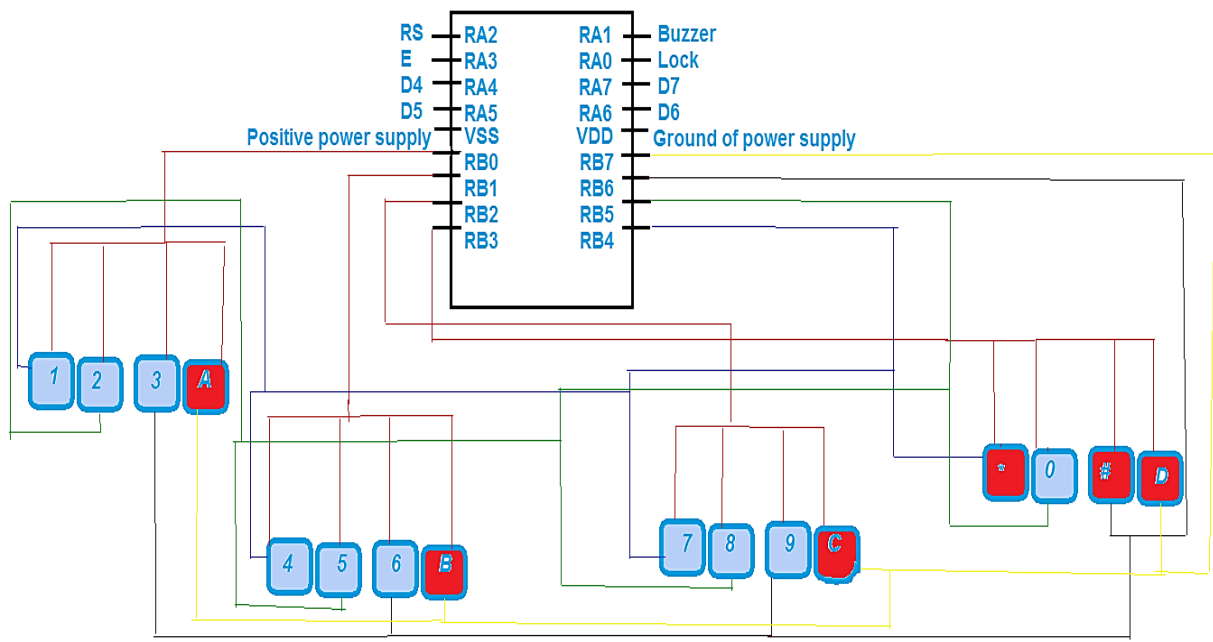

*Figure 1: Microcontroller pins assignment*

*B. LCD Implementation*

The other main part of the system implementation is LCD implementation. The LCD configures to operate in 4 bits interface mode, because of PIC16F628A ports limitations. Sending the command 0x20 then 0x28 will configure the LCD to work in 4 bits interface mode with two lines[4], which means that the maximum characters can be displayed are 40 characters for a 20x2 LCD used by the system.

The function *send_word* is responsible for the implementation of sending a word to LCD in a 4 bit interface mode. Following steps are performed by the *send_word* function to send a single word in 4 bit interface mode [4]:

1. Set the E line low.
2. Set the RS line high for sending data, or low for sending a command.
3. Set the E line high.
4. Put the four most significant bits on DB4 to DB7.
5. Set the E line low.
6. Set the E line high.
7. Put the four least significant bits on DB4 to DB7.
8. Set the E line low The E line should remain low until clocking in another byte.

## IV. SYSTEM OPERATION

When the system starts, it enters directly into the main loop. In its entire operation time, it is interchanging only between two states. First state continuously scans the input key and the other state is go to sleep on PORTB, because PORTB is configured to PULL UP which makes the system to operate in power saving mode. This saves battery power and improves the system operation lifetime, since the microcontroller doesn't require execution of instructions continuously even if nobody interacts with the system.

The system moves from sleep to wake up state while pressing any key, which change the status of the PORTB. PIC 16F628a has the PULL UP on PORTB property for all the PROTB pins [3].

When the system in a lock state, and user tries to open the lock, the user have to enter ten digits that must conform to the ten digits stored in the EEPROM. The following figure prompts to the user to enter the password:

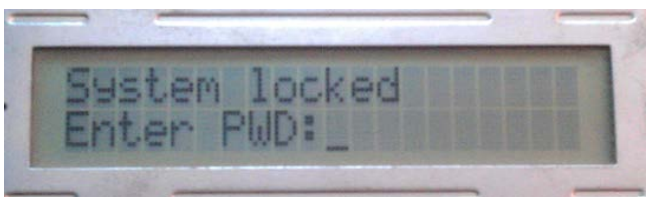

*Figure2: Message prompt to user to enter the password*

Then the user enters the password which is limited to 10 digits and can combined with the special characters '*' and '#' making the password more complex to be broken.

The following is the figure showing the user entering the password to the system (digits displayed as '*' in LCD but they are displayed here as for clarity to the readers).

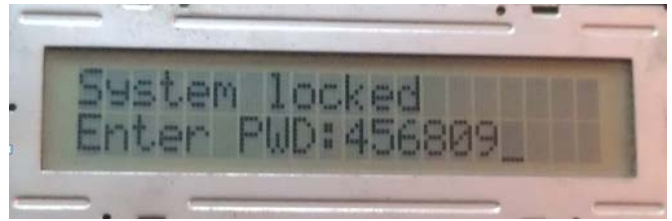

*Figure3: User entered the password to open the lock*

After entering the password, user is required to press the *Enter* button. Then the function *verify_password* would be called to verify if the input password is similar to the one that is stored in the EEPROM. If the two passwords are same, then the lock will open and the message "verify successfully" would be displayed on LCD.

Following are the steps for *verify_password,* function which would make a comparison between two arrays if they have the same value or not. Indirect memory access is used for the comparison process [2].

1. Initialize count with value 10.
2. Make FSR point to start address of the first array.
3. Store the value of location that FSR point to it in variable name *data1*(first array).
4. Increment FSR and save its value in *save1* variable.
5. Test if the function in the first step in the loop.
6. If yes will make FSR point to start address of the second array and go to step 8.
7. If no will put the value of variable *save2* to FSR.
8. Store the value of location that FSR point to it in variable name *data2* (second array).
9. Increment FSR and save the value of it in *save2*.
10. Put the value of *save1* in FSR.
11. If *data1* equal *data2* then:
✓ Decrement count.
✓ If count equal zero will go to step 13.
✓ Else will go to step 3.
12. Else go to out of loop and assign false to flag.
13. Return and assign true to flag.

The next function is *ReadEEPROM* which is used to read the stored password from to EEPROM to the file register in array to further compare with the password entered by the user.

Following are the steps performed by *ReadEEPROM:*

1. Initialize count with value 10.
2. Assign start address of EEPROM to specific pointer.
3. Make FSR point to start address of array, which will contain the value of EEPROM.
4. Increment FSR.
5. Read value of one location from EEPROM and write it in the array.
6. Increment the pointer that point to EEPROM.
7. Decrement count. If equal to zero, it mean the EEPROM password has been read completely then return. Otherwise, go to step 4.

The other important function is *WriteEEPROM* function, which writes the new password to EEPROM when the user wants to change the password. Writing to the EEPROM is critical operation which should not be performed accidentally [3], because the results would be permanent, unless and until we re-write the same location of EEPROM. For this reason, two values *AAh* and *55h* must be written to EECON2 register one after another to be sure the program want to really write to EEPROM.

Following are the steps performed by the function *WriteEEPROM*:

1. Initialize count with value 10.
2. Assign start address of EEPROM to specific pointer.
3. Make FSR point to start address of array which contain the value that must be store in EEPROM
4. Increment FSR.
5. Read value of one location from array to temporary variable.
6. Write the value 0xAA to EECON2.
7. Write the value 0x55 to EECON2.
8. Write the temporary variable to EEPROM.
9. Increment the pointer that point to EEPROM.
10. Decrement count if equal zero that meaning all password has finished write in all location of EEPROM and return, otherwise will go to step *4*

The system has the buzzer to represent an alarm when the user enters a password three times the buzzer will give a high sound. This pin could be configured with rely to switch a high power alarm when a security breach occurs.

The proposed system is very simple and has very few components involved, because all the functions are implemented in the firmware that is installed in the PIC microcontroller.

Figure 4, shows us the complete hardware components used in the system on the breadboard.

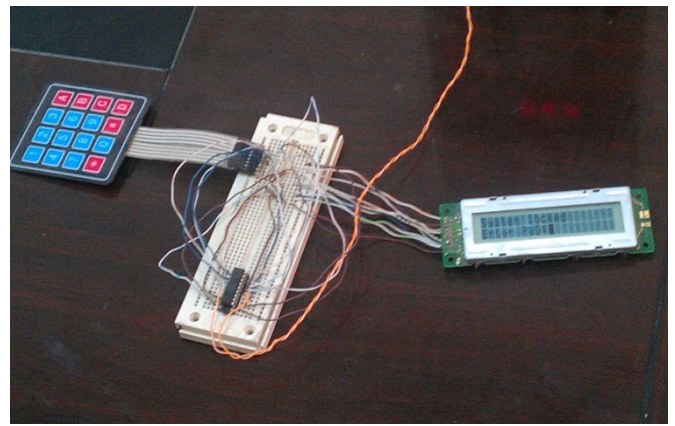

*Figure 4: Complete hardware components of the system*

It is evident from the figure, that there are only a few electronic components involved in the working of this system. This is really a great benefit of using an embedded system, which makes the hardware architecture very simple, but the complexity ultimately moves to the firmware itself according to the concepts of software and hardware boundary.

## V. CONCLUSION

Programming the LCD with 4 bits interface mode is very efficient and the delay of sending a single byte in two steps was not noticeable, irrespective of reserving the 4 bins port in a limited ports microcontroller.

The other issue is using PULL UP on PORTB, which proved beneficial for the system, because it puts the system in the sleep mode within the infinite loop while waiting for a key press. So when there is no interaction with the system, it saves batteries power and improves the functional life of the system. Hence there is no wastage of CPU cycles for unimportant functions.

After implementation of the system using real hardware components, there was one technical difficulty I encountered, and want to emphasize in this paper, was the implementation of keypad. As many researches state that, Columns has to be set as input to be read, to check their status to know which key has been pressed, and the rows as output, which can be set by firmware to 1 or 0 [5,6]. This approach did not work here because, if 4 pins are set to input simultaneously their state would be influenced when one of them is changed causing inaccurate key scan for reading the pressed key. The solution found was to set every port pin of column to input then output every time we check the column status. In this case, only one pin would be set as input at a time and the others are set to output, which would not be influenced by changing state of that pin. This gives us an accurate and smooth keypad operation.

The system is intact and sound, and with some improvements and operational testing it can be considered as a successful product and can be shipped to the market.